\documentclass{article}

\usepackage[round]{natbib}
\usepackage{arxiv}
\usepackage{caption}
\usepackage{amsmath}

\usepackage[utf8]{inputenc} 
\usepackage[T1]{fontenc}    
\usepackage{url}            
\usepackage{booktabs}       
\usepackage{amsfonts}       
\usepackage{nicefrac}       
\usepackage{microtype}      
\usepackage{lipsum}
\usepackage{graphicx}
\graphicspath{ {./images/} }

\title{Efficient Fine-tuning Methodology of Text Embedding Models for Information Retrieval: Contrastive Learning Penalty (CLP)}

\author{
 Jeongsu Yu \\
  Department of Convergence Security\\
  Kangwon National University\\
  \texttt{yjshi25@kangwon.ac.kr} \\
}

\begin{document}
\maketitle
\begin{abstract}
Text embedding models play a crucial role in natural language processing, particularly in information retrieval, and their importance is further highlighted with the recent utilization of RAG (Retrieval-Augmented Generation). This study presents an efficient fine-tuning methodology encompassing data selection, loss function, and model architecture to enhance the information retrieval performance of pre-trained text embedding models. In particular, this study proposes a novel Contrastive Learning Penalty function that overcomes the limitations of existing Contrastive Learning. The proposed methodology achieves significant performance improvements over existing methods in document retrieval tasks. This study is expected to contribute to improving the performance of information retrieval systems through fine-tuning of text embedding models. The code for this study can be found at https://github.com/CreaLabs/Enhanced-BGE-M3-with-CLP-and-MoE, and the best-performing model can be found at https://huggingface.co/CreaLabs.

\end{abstract}


\section{Introduction}
Text embedding models play a crucial role in natural language processing, particularly in information retrieval, by mapping text data into a semantically rich vector space. The importance of information retrieval has been further highlighted with the recent utilization of RAG (Retrieval-Augmented Generation) \citep{DBLP:journals/corr/abs-2005-11401} to address the issues of hallucination and outdated information in large language models (LLMs).  Pre-trained text embedding models on a massive corpus have significantly improved the quality of text representation. BGE M3-Embedding \citep{chen2024bgem3embeddingmultilingualmultifunctionality} is a representative model that shows outstanding performance in multilingual text embedding and information retrieval.

This study proposes an efficient fine-tuning methodology to enhance the information retrieval performance of pre-trained text embedding models by specializing them to a specific domain:

1. Efficient Training Data Selection Technique:  Applies ANCE (Approximate Nearest Neighbor Negative Contrastive Estimation) \citep{xiong2020approximatenearestneighbornegative} for selecting negative samples in the training data.

2. Contrastive Learning Penalty (CLP): Analyzes the limitations of existing Contrastive Learning functions and proposes a novel loss function to overcome them.

3. Mixture of Experts (MoE) Technique \citep{gupta2022sparselyactivatedmixtureofexpertsrobust}:  Generates optimized embeddings based on the characteristics of the input text to effectively respond to various types of queries.

To validate the effectiveness of the proposed methodology, we evaluated its performance on document retrieval tasks in three languages using the nDCG metric. The experimental results demonstrate that the proposed methodology achieves significant performance improvements compared to existing methods.

\section{Related Work}
In the past, keyword-matching based retrieval methods like BM25 were mainly used for document retrieval. BM25 calculates a similarity score considering term frequency, inverse document frequency, and document length \citep{robertson1995okapi, crestani1998is}. Although BM25 is an efficient and widely used method, it has the following limitations:

1. Vocabulary mismatch: If different words with the same meaning are used in the query and the document, BM25 may not accurately capture the relevance. For example, BM25 recognizes "car" and "vehicle" as different words even though they have the same meaning.

2. Difficulty in understanding meaning: Since BM25 only considers the frequency and distribution of words, it does not capture the meaning or context of words. Therefore, it may have difficulty accurately matching the intent of the query with the content of the document.

3. Ignoring syntactic information: BM25 does not consider syntactic information such as word order or sentence structure. Therefore, "how to get from Seoul to Busan" and "how to get from Busan to Seoul" can be treated the same in BM25.

These limitations can be overcome with the advent of text encoders, i.e., text embedding models, from pre-trained language models \citep{reimers2019sentencebertsentenceembeddingsusing, ni2021sentencet5scalablesentenceencoders}. 
\paragraph{Dense Retrieval.} Dense Retrieval (DR) using text embedding models enables search results that understand the meaning of text, not just word matching. DR maps text to a high-dimensional vector space using deep learning and performs search by measuring similarity based on distances in this vector space \citep{karpukhin2020densepassageretrievalopendomain}.

\paragraph{Contrastive Learning.} The learning of text embedding models for DR has been advanced by the advent of Contrastive Learning techniques \citep{hadsell2006dimensionality, chen2020simpleframeworkcontrastivelearning, gao2022simcsesimplecontrastivelearning}. Contrastive Learning (CL) aims to learn effective representations by pulling together semantically close neighbors and pushing away irrelevant ones. This assumes a dataset $D=(q _{i} ,p _{i}^{+} ,P _{i}^{\prime} ) _{i=1}^{m}$ composed of semantically related pairs, where $q _{i}$ denotes a query, $p _{i}^{+}$ denotes a positive sample for $q _{i}$ and $P _{i}^{\prime}$ denotes a set of negative samples for $q _{i}$. Following the Contrastive Learning framework of \citet{chen2020simpleframeworkcontrastivelearning}, we use a cross-entropy objective function with negative samples from within a mini-batch. Specifically, when $h_i$, $h^-_i$,$P _{i}^{\prime}$ are the representations of $q _{i}$, $p _{i}^{+}$, $P _{i}^{\prime}$ respectively, the training objective function on a mini-batch of $N$ pairs is:

\begin{equation}
LITER_{i} = -\log \frac{\exp(\text{sim}(h_{i}, h_{i}^{+}) / \tau)}{\sum_{H_{i} \in \{h_{i}^{+}, H^{\prime}\}}^{N} \exp(\text{sim}(h_{i}, H_{i}) / \tau)} 
\end{equation}

where $r$ is a temperature hyperparameter and $sim(h_1,h_2)$ is the cosine similarity  $\frac{h_1^T h_2}{||h_1|| \cdot ||h_2||}$.

To perform CL, we need to construct pairs $D=(q _{i} ,p _{i}^{+} ,P _{i}^{\prime} ) _{i=1}^{m}$. In this process, effective negative sampling has a significant impact on learning performance. The ANCE  \citep{xiong2020approximatenearestneighbornegative} study revealed that using uninformative negative samples that are distant from the query leads to problems such as decreased gradient values, increased stochastic gradient variance, and slow learning convergence. Therefore, utilizing informative negative samples that are closer to the query is more effective.

\section{Fine-tuning Methods}

\subsection{Contrastive Learning Penalty (CLP)}
CL simply minimizes the distance between $q_i$ and its semantically related counterpart $p^+_i$, while maximizing the distance to the unrelated $p^-_i$. However, this learning method does not consider the distance between $H^-_i$ and its semantically related queries, $Q^*_i$, where $Q^*_i$ is the set of positive queries for each document in $H^-_i$. For efficient learning, while minimizing the distance between $q_i$ and $p^+_i$ and maximizing the distance to $H^-_i$, we need to ensure that this does not negatively affect the distance between $H^-_i$ and $Q^*_i$. In this study, we propose a novel CLP to address the negative impact on the distance between $H^-_i$ and $Q^*_i$ in existing CL.

CLP imposes a penalty that increases as the distance between $H^-_i$ and $Q^*_i$ grows in the existing Contrastive Learning Loss function. The specific formula is as follows:

\begin{equation}
\text{LITER}_i = (\lambda - 1) \log \frac{\exp(\text{sim}(h_i, h_i^+) / \tau)}{\sum_{H_i \in \{h_i^+, H^\prime\}}^N \exp(\text{sim}(h_i, H_i) / \tau)} + \lambda \left( 1 - \sum_{h^* \in H^*} \text{sim}(h^\prime, h^*) \right)
\end{equation}

where $\lambda$ is the weight of the penalty, and $H^*$ represents the representations of $Q^*_i$.

\begin{figure}[h]
    \centering
    \includegraphics[width=1\textwidth]{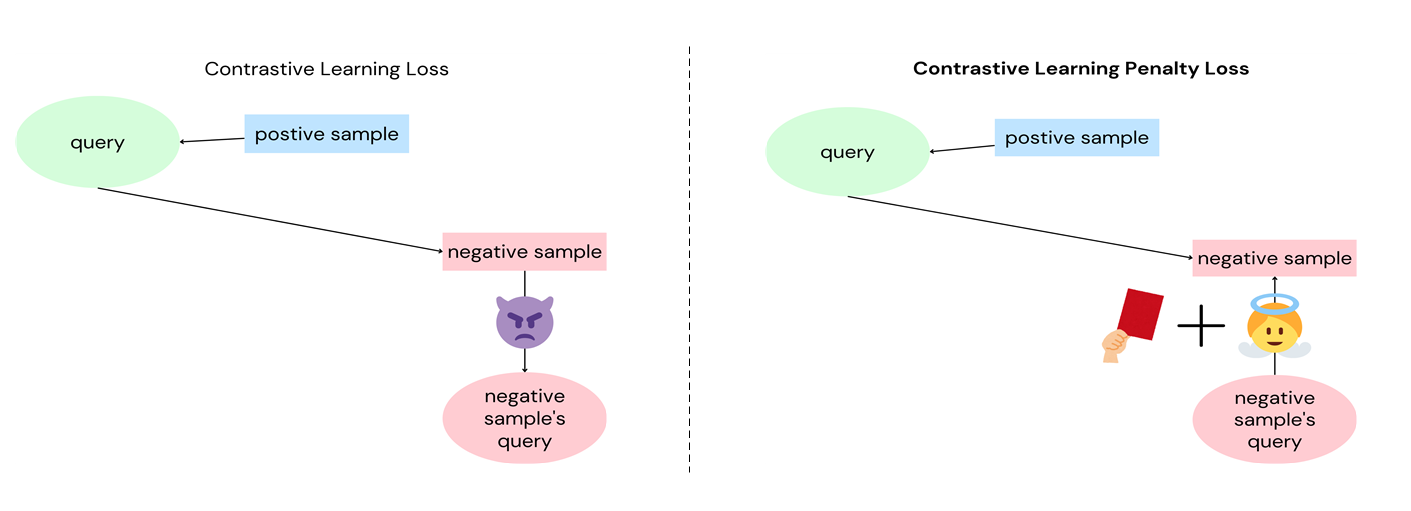} 
    \caption{The difference between CL and CLP}
\end{figure}

\subsection{Negative Sampling}
In this study, we utilize the top 10 most similar samples from the dense retrieval results of the training corpus of the model being trained, excluding the positive sample, as negative samples, referring to the ANCE methodology. This provides information-rich samples to induce effective learning and prevents problems caused by using uninformative negative samples \citep{xiong2020approximatenearestneighbornegative}.

\subsection{Mixture of Experts(MoE)}
Existing methodologies utilized dense networks that use the same weights for all tasks, regardless of input characteristics. However, this caused a problem where inputs with different characteristics interfered with each other in model training. To address this issue, Mixture of Experts (MoE) was introduced. MoE shares only some weights and trains the remaining weights according to the input characteristics. By assigning tasks to suitable experts for the input through a task-aware gating function that analyzes input characteristics, it improves transfer learning to low-resource tasks, enables efficient generalization, and prevents capacity loss in existing models \citep{gupta2022sparselyactivatedmixtureofexpertsrobust}.

This study leverages the advantages of MoE by applying the MoE structure to the intermediate layer of the existing text embedding model. We froze and trained the remaining parameters excluding the MoE layer.

\section{Experiment}
\subsection{Evaluation Setup}

\subsubsection{Metric}

In this study, we fine-tuned the BGE M3-Embedding model for specific languages (Korean, Hindi, and Persian) using the multilingual document retrieval dataset MIRACL (Table \ref{table:MIRACL}) and evaluated its performance. Korean is an agglutinative language, Hindi is an inflectional language, and Persian belongs to the Indo-European language family. By selecting these languages, we aimed to comprehensively analyze various language types.

\begin{table}[h]
    \centering
    \includegraphics[width=1\textwidth]{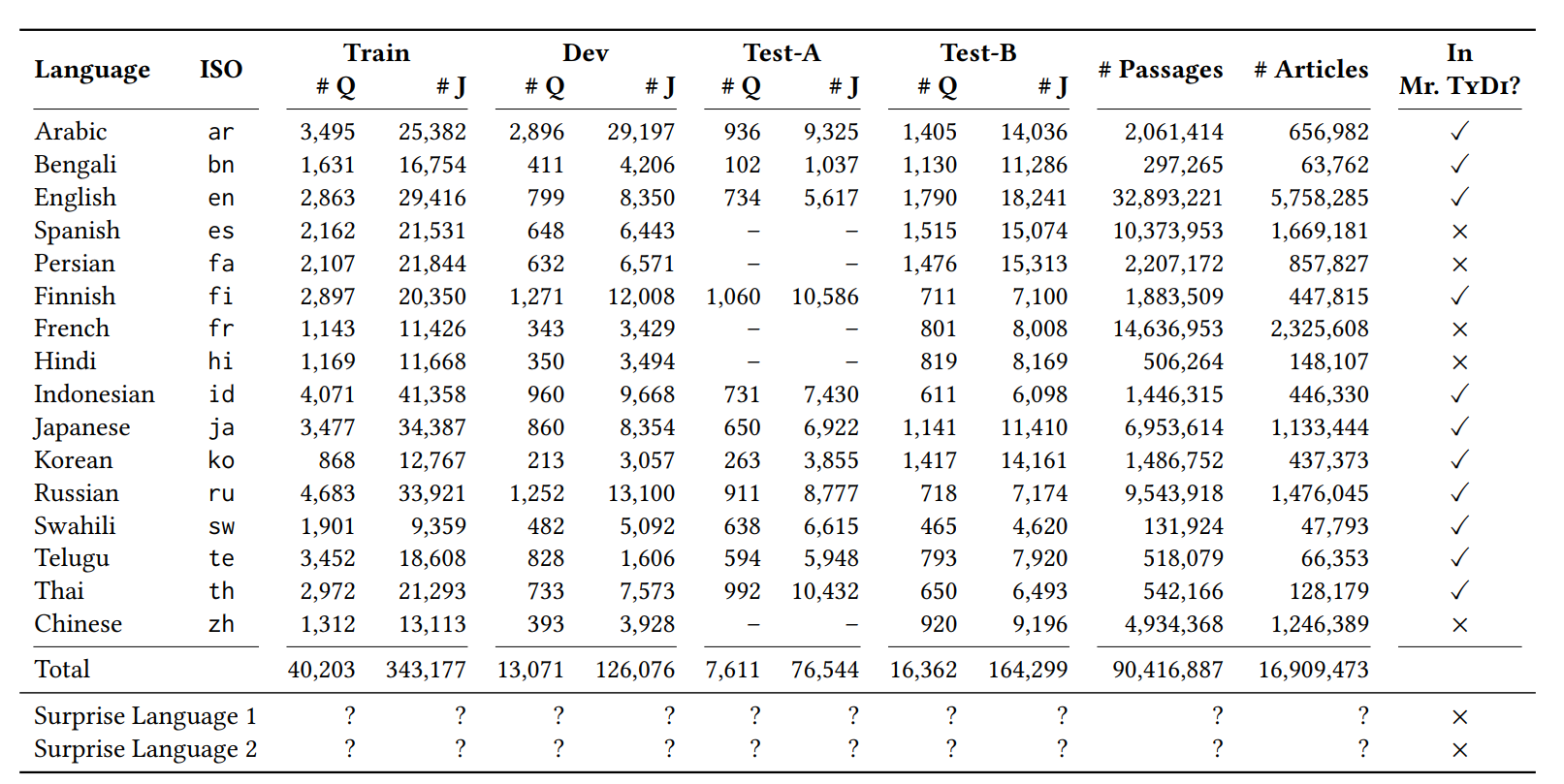} 
    \vspace{1pt}
    \caption{MIRACL dataset \citep{zhang2022makingmiraclmultilingualinformation}}
    \label{table:MIRACL}
\end{table}

The dense retrieval results of the fine-tuned models are evaluated using the nDCG metric. nDCG is an evaluation metric that considers the ranking of search results, assigning higher scores when more relevant documents are ranked higher, and is widely used in the field of information retrieval. The BGE M3-Embedding model achieved higher nDCG scores than previous models on MIRACL (Table \ref{table:BGE_M3_MIRACL}).

\begin{table}[h]
    \centering
    \includegraphics[width=1\textwidth]{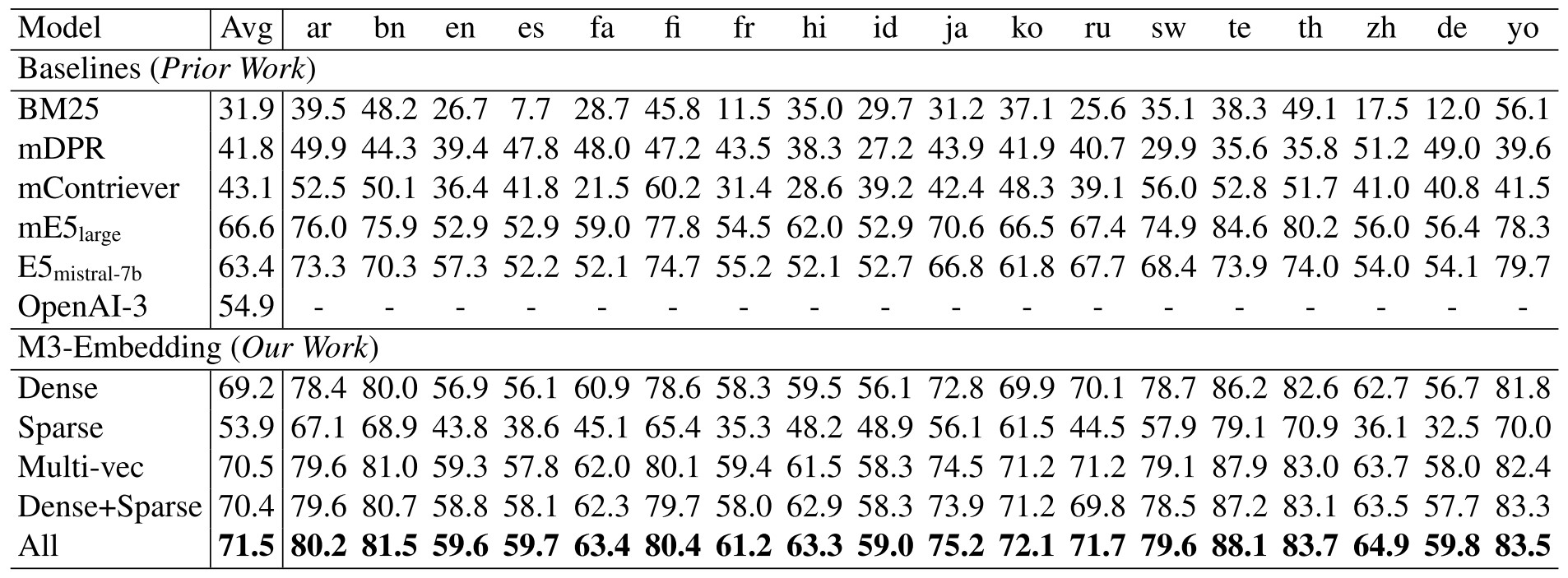} 
    \vspace{1pt}
    \caption{BGE M3-Embedding on MIRACL dev set (nDCG@10)
 \citep{chen2024bgem3embeddingmultilingualmultifunctionality}}
    \label{table:BGE_M3_MIRACL}
\end{table}

\subsubsection{Training details}

This model was trained on a single 3090 GPU with the following hyperparameters:

\begin{itemize}
\item Learning rate: 1e-5
\item Mixed precision training (fp16) 
\item Number of training epochs: Table \ref{table:epoch}
\item Per-device training batch size: 1
\item Gradient accumulation steps: 4
\end{itemize}

For CLP, the penalty weight was set to 0.1. For the MoE layers, the number of experts was set to 2 and the number of experts per token was set to 1.

In this study, we applied four methods based on the BGE M3-Embedding model to improve its performance. Detailed explanations of each method are provided in the following paragraphs. To find the model with the optimal performance, we fixed the hyperparameters for each method and trained the model while only changing the epoch value. The epoch that showed the best performance for each method is as follows:

\begin{table}[h]
  \centering
  \begin{tabular*}{0.8\textwidth}{@{\extracolsep{\fill}}lccc}
    \hline
    method/Number of training epochs & ko & fa & hi \\
    \hline
    random dataset    & 1   & 1 & 1 \\
    ANCE dataset     & 2   & 1 & 1  \\
    ANCE-CLP   & 2   & 1 & 1 \\
    ANCE-CLP-intermediate   & 2   & 1 & 1  \\
    ANCE-CLP-moe-intermediate  & 3   & 3 & 3  \\
    \hline
  \end{tabular*}
  \vspace{10pt}
  \caption{Optimal epoch value for each method} 
  \label{table:epoch}
\end{table}

\subsection{Performance Comparison}

\begin{table}[h]
  \centering
  \begin{tabular*}{0.8\textwidth}{@{\extracolsep{\fill}}lcccc}
    \hline
    method/language & ko & fa & hi & average \\
    \hline
    baseline(BGE M3-Embedding)    & 64.36   & 51.13 & 52.35 & 55.95 \\
    random dataset     & 59.61  & 47.38 & 51.78 & 52.92 \\
    ANCE dataset   & 66.89  & 50.88 & 54.53 & 57.43 \\
    ANCE-CLP  & 67.05 & 52.39 & 54.75 & 58.06 \\
    ANCE-CLP-intermediate  & 66.92 & 55.1 & 56.34 & 59.45 \\
    ANCE-CLP-moe-intermediate  & \textbf{67.74} & \textbf{55.56} & \textbf{56.36} & \textbf{59.89} \\ 
    \hline
  \end{tabular*}
  \vspace{10pt}
  \caption{MIRACL dev set (measured by nDCG@5)} 
  \label{table:result}
\end{table}

\subsubsection{Train Dataset}

This study compared and analyzed negative sampling techniques in CL. Comparing the conventional method of random negative sampling with the ANCE methodology, the following performance differences were observed:

\begin{itemize}
\item random dataset: This method employed random negative sampling. This approach resulted in lower performance (52.92) compared to the baseline model (55.95). This is likely because random sampling has a higher chance of selecting negative samples with low information content, making it less effective for model training.
\item ANCE dataset: This method utilized the ANCE methodology for negative sampling. This approach achieved higher performance (57.43) than the baseline model (55.95). It is analyzed that ANCE contributed to the performance improvement by effectively selecting hard negative samples, which provide useful information for model learning.
\end{itemize}

These experimental results show that the negative sampling technique in CL can significantly influence model performance. In particular, it was confirmed that a technique that effectively selects hard negative samples, such as ANCE, is effective in improving the performance of the model.

\subsubsection{Loss Funtion}

This study compared and analyzed the performance of the proposed CLP and existing CL. Applying CLP requires positive queries corresponding to negative samples, but the MIRACL dataset used in the experiment does not include this information.

Therefore, we utilized Gemini 1.5 Pro to generate synthetic data for positive queries. Specifically, we generated negative sample's positive query through the Gemini 1.5 Pro model using the prompt in Table \ref{table:prompt}.

\begin{table}[h]
    \centering
    \includegraphics[width=1\textwidth]{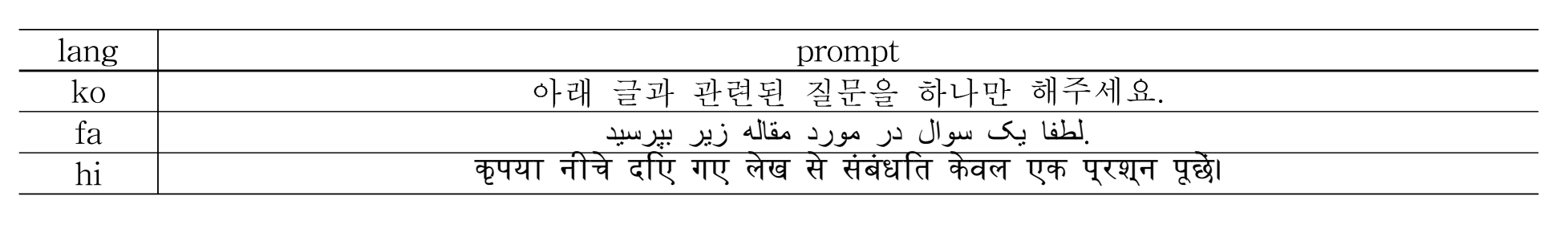} 
  \caption{MIRACL dev set (measured by nDCG@5)} 
  \label{table:prompt}
\end{table}

As shown in Table \ref{table:result}, the experimental results demonstrate a performance improvement when applying CLP (ANCE-CLP) with a score of 58.06, compared to the existing CL (ANCE dataset) which scored 57.43. Notably, in the case of Persian, applying the existing CL resulted in a performance decrease (50.88) compared to the baseline model (51.13). However, applying CLP showed a significant performance improvement (52.39), demonstrating the superiority of CLP.

However, CLP took 3 minutes and 16 seconds to train one epoch on the MIRACL Korean learning dataset (854 instances), while the existing CL took 4 minutes and 18 seconds. The longer training time for CLP is attributed to the additional computational step of calculating the similarity between the negative sample and the negative sample's query.

\subsubsection{Model Architecture}
To determine the optimal location for applying Mixture of Experts(MoE), we conducted an experiment where only the intermediate layer of the model was trained, while keeping other layers frozen. Interestingly, training only the intermediate layer (ANCE-CLP-intermediate) resulted in higher performance (59.45 -> 59.89) than training all parameters of the model (ANCE-CLP). In this experiment, the intermediate layer serves to expand the dimension from 1024 to 4096.

Based on these experimental results, we applied MoE to the intermediate layer (ANCE-CLP-moe-intermediate) while freezing other parameters. This ultimately yielded the best performance of 59.89.

However, applying MoE has the drawback of increased inference time. Embedding the MIRACL Korean corpus dataset (371,688 instances) took 3 hours and 7 minutes for the baseline model, while the MoE model required 5 hours and 1 minute.

\begin{figure}[h]
    \centering
    \includegraphics[width=0.7\textwidth]{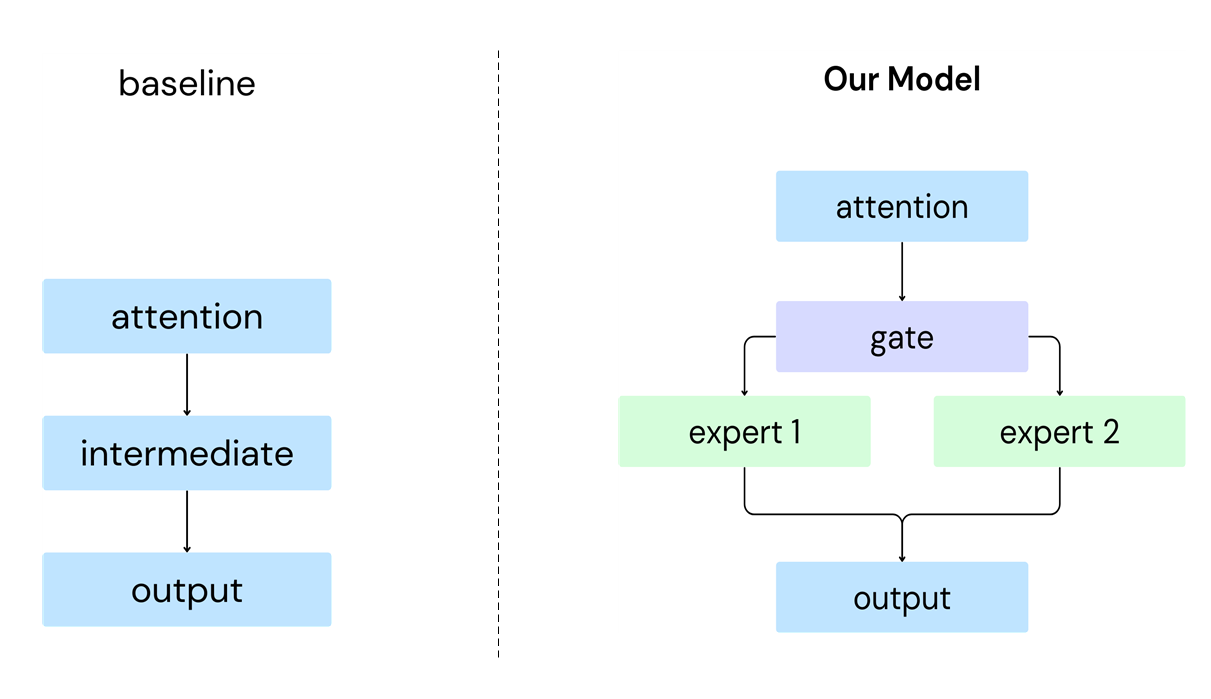} 
    \caption{Model Architecture}
\end{figure}

\section{Conclusion}

This study proposes an efficient fine-tuning methodology for specializing pre-trained text embedding models to a specific domain, thereby enhancing information retrieval performance. We aimed to maximize model performance throughout the entire fine-tuning process, from training data construction to loss function and model structure.

Our key improvements include:

1.  Efficient Training Data Selection: We utilized ANCE to select informative negative samples, leading to more effective model learning.

2.  Contrastive Learning Penalty (CLP): We introduced a novel loss function, CLP, which addresses limitations of existing contrastive learning by considering the relationship between negative samples and their corresponding positive queries.

3.  Mixture of Experts (MoE) Application: We applied MoE to the intermediate layer of the text embedding model to generate optimized embeddings tailored to diverse input text characteristics.

Experiments conducted on a multilingual document retrieval dataset (MIRACL) across Korean, Hindi, and Persian demonstrated the effectiveness of the proposed methodology. By applying all proposed methods, we achieved a final performance improvement of approximately 5 points compared to the baseline.

The main contributions of this study are as follows:

\begin{itemize}
\item We demonstrate the effectiveness of combining ANCE, CLP, and MoE for fine-tuning pre-trained text embedding models in information retrieval tasks.
\item The newly proposed CLP significantly improved performance, especially for Persian, where conventional CL exhibited performance degradation.
\end{itemize}

This study presents a novel approach to fine-tuning text embedding models and is expected to contribute to improving the performance of information retrieval systems. Future research will focus on validating the applicability of our method to various languages and domains and further enhancing the efficiency of CLP.

The code for this study is available at https://github.com/CreaLabs/Enhanced-BGE-M3-with-CLP-and-MoE, and the final model can be found at https://huggingface.co/CreaLabs.

\bibliography{references}

\end{document}